\documentclass[12pt]{article}
\usepackage{amsmath}
\usepackage{amssymb}

\usepackage{subfigure}
\usepackage{psfrag}
\usepackage{epsfig}
\usepackage{graphicx}
\usepackage{tikz}
\usepackage{cite}
\usepackage{hyperref}

\newcount\xrefpos \xrefpos=0
\newcount\yrefpos \yrefpos=0
\newcount\xput \xput=0
\newcount\yput \yput=0

\def\refpos#1 #2 #3{\global\xrefpos=#1 \global\yrefpos=#2
                         \rlap{$\smash{#3}$}}
\def\put #1 #2 #3{\xput=#1 \yput=#2
                  \advance\xput by -\xrefpos
                  \advance\yput by -\yrefpos
                  \rlap{\kern\the\xput truebp
                        \vbox to 0pt{\vss\hbox{$\displaystyle #3$}
                        \kern\the\yput truebp}}}
\def\beginlabels\refpos#1\endlabels{\hbox{$\refpos#1$}}

\newcommand{\ba}{\begin{eqnarray}}
\newcommand{\ea}{\end{eqnarray}}

\usepackage{amsfonts,amssymb}

\begin{document}

 \begin{center}
  {\Large \bf Comments on Non-Fermi Liquids in the Presence of a Condensate}

\bigskip
\bigskip
\bigskip
\bigskip

\vspace{3mm}

Pallab Basu$^{a,}$\footnote{email: pallab@phas.ubc.ca} , Jianyang He $^{a,}$\footnote{email: jyhe@phas.ubc.ca}, Anindya Mukherjee$^{a,}$\footnote{email: anindya@phas.ubc.ca}, Moshe Rozali $^{a,}$\footnote{email: rozali@phas.ubc.ca} and Hsien-Hang Shieh$^{a,b.}$\footnote{email: shieh@phas.ubc.ca}

\bigskip\medskip
\centerline{$^a$\it Department of Physics and Astronomy}
\smallskip\centerline{\it University of British Columbia}
\smallskip\centerline{\it Vancouver, BC V6T 1Z1, Canada}

\bigskip \centerline{$^b$\it Perimeter Institute for theoretical Physics}
\smallskip\centerline{\it 31 Caroline Street North}
\smallskip\centerline{\it Waterloo, Ontario }
\smallskip\centerline{\it Canada N2L 2Y5}
 \end{center}

 \bigskip\bigskip\bigskip

% \maketitle

%\title{Comments on Non-Fermi Liquids in the Presence of a Condensate}
%\author{Pallab Basu, Jianyang He, Anindya Mukherjee and Moshe Rozali}

% Label conventions:
% 	equation labels begin with 	eq:
% 	section labels begin with 	sec:
% 	figure labels begin with 	fig:

\abstract{We study the effects of a scalar condensate on a class of 2+1 dimensional non-Fermi liquids by introducing fermionic probes in the corresponding asymptotically $AdS_4$ black hole backgrounds. For  the range of parameters and couplings we consider we find gapless fermionic excitations whose properties are model-dependent.}

%\tableofcontents
\newpage
\section{Introduction and Conclusions}
\label{sec:introduction}

An important problem in condensed matter physics is the understanding of fermionic systems at strong coupling. The strongly coupled nature of these systems, and numerical instabilities due to the fermion sign problem make it hard to investigate such systems using conventional analytic or numerical methods. A non-Fermi liquid picture is believed to emerge in some situations; in these cases the fermionic system has a well-defined Fermi surface but differ significantly in other respects from the Landau Fermi liquid. Non-Fermi liquids are interesting for a number of reasons. For example, it has been proposed that the normal state of high-$T_c$ superconductors, and metals close to a quantum critical points are examples of non-Fermi liquids \cite{Anderson:1990,Varma:1989}.

In string theory, one can use the gauge-gravity duality to investigate various strongly coupled systems, which map a strongly coupled field theory to a weakly coupled gravitational system \cite{Maldacena:1997re}. Recently there has been much progress in finding gravitational analogues of condensed matter and fluid dynamics phenomena, see for example \cite{Hartnoll:2009sz,Herzog:2009xv,Rangamani:2009xk, Sachdev:2008p11034, Horowitz:2010gk} and references therein.

An interesting recent development is the work by Liu et al \cite{Liu:2009dm}, where the authors have studied probe fermions in an extremal $AdS$ black hole geometry\footnote{For other work on holographic non-Fermi liquids see \cite{Lee:2008xf,Iqbal:2009fd, Iqbal:2009fd, Cubrovic:2009ye,Rey:2008zz, Basu:2009qz}.}. The authors calculate the retarded Green function of the fermionic field, showing that the system behaves like a non-Fermi liquid, or in other words in a certain range of parameters a well-defined Fermi surface seems to exist. Some interesting non-Fermi scaling behavior can be demonstrated \cite{Liu:2009dm, Faulkner:2009wj}.

The main motivation of this note is to study of the effects of  a scalar condensate on the fermionic system\footnote{For previous work on such effects see \cite{Faulkner:2009wj, Chen:2009p10965, Gubser:2009p10980,Gubser:2009p10962, Albash:2009wz, Albash:2010p11042}.}. To this end, we extend the black hole background by turning on a scalar condensate; In this paper we will mostly be concerned with the non-extremal geometries. Since the condensate strength can become large, it becomes necessary to consider its effects on the gravitational background.  We next introduce probe fermions in this geometry with a suitable coupling to the scalar. The coupling to the scalar, and other parameters we consider here (such as the fermion charge and the temperature) extend and generalize previous work \cite{Faulkner:2009wj}. We find  that the system always contains sharp fermionic excitations, which are gappless and become stable in the zero temperature limit. The detailed properties of those fermionic excitations, depend strongly on the parameters chosen, in a way we discuss in detail. The robustness of the Fermi surface is perhaps surprising, and might be related to the large $N$ limit.

The plan of this note is as follows. We describe the required background in sections 2 and 3. Section 2 is devoted to introducing the models of holographic superconductivity we will use here. Section 3 introduces the tools needed to probe those systems using bulk fermions. We then introduce features of the Fermi surface (as exhibited by properties of the fermion spectral functions) is sections 4 and 5: in section 4 we discuss the high temperature case, and in section 4 we discuss the low temperature (condensed) phase.

In this note we concentrated on the physics of the holographic superconductors obtained from the simplest Abelian Higgs model in the bulk \cite{Gubser:2005p11024, Gubser:2008p10986,Hartnoll:2008vx, Hartnoll:2008kx}. There is another set of models utilizing non-Abelian gauge fields in the bulk \cite{Gubser:2008p10985}, which are more interesting in some ways. For instance,  the bulk couplings are more constrained, and the model allows for both s-wave and p-wave \cite{Gubser:2008p10984} superconductors\footnote{For recent work on this set of models see \cite{Basu:2009p10975, Ammon:2009p10970}.}. We hope to report on our results studying this set of models shortly \cite{toappear}.

\section{Asymptotically $AdS_4$  Black Holes}
\label{sec:adsbh}

The set of backgrounds we are interested in asymptote to an $AdS_4$ geometry, with possibly some profiles for the gauge and scalar fields.  This corresponds to a 2+1 dimensional dual field theory at a finite temperature, with a finite chemical potential, and possibly resulting in a condensation of a scalar operator. We are interested in various such geometries, which dominate the thermodynamics of the dual field theory at different temperature ranges.

The system is described by gravity coupled to a Maxwell field and a charged scalar field (an Abelian Higgs model), with the Lagrangian:
\begin{equation}
\label{eq:lagrangian}
\mathcal{L} = R + \frac{6}{L^2} - \frac14 F^{\mu\nu}F_{\mu\nu} - |\nabla\psi - iq_{b}A\psi|^2 - V(|\psi|),
\end{equation}
where $F$ is the electromagnetic field strength, and $q_{b}$ is the charge of the scalar field $\psi$.  We follow the conventions in \cite{Hartnoll:2008kx}.

Assuming solutions with spherical symmetry, the metric has the general form:
\begin{equation}
\label{eq:metric}
ds^2 = -g(r)e^{-\chi(r)}d t^2 + \frac{d r^2}{g(r)} + r^2(dx^2+dy^2)
\end{equation}
with $A = \phi(r) dt$ and $\psi = \psi(r)$. By a suitable gauge choice we can assume $\psi$ to be real, and the scaling symmetries of the metric allow us to set $L = 1$. The full non-linear equations of motion are given by:
\begin{eqnarray}
\label{eq:eomstart}
& \psi'' + \left(\frac{g'}{g} - \frac{\chi'}{2} + \frac2r \right)\psi' + \frac{q_{b}^2 \phi^2 e^\chi}{g^2}\psi - \frac{V'(\psi)}{2g} = 0 & \nonumber\\
& \phi'' + \left(\frac{\chi'}2 + \frac2r\right)\phi' - \frac{2q_{b}^2\psi^2}{g}\phi = 0 & \nonumber\\
& \chi' + r\psi'^2 + \frac{rq_{b}^2 \phi^2 \psi^2 e^\chi}{g^2} = 0 & \nonumber\\
\label{eq:eomend}
& g' + \left(\frac1r - \frac{\chi'}2 \right)g + \frac{r \phi'^2 e^\chi}4 - 3r + \frac{rV(\psi)}2 = 0, &
\end{eqnarray}
with the simple potential $V(\psi) = \frac12 m_{\psi}^2 \psi^2$ where $m_{\psi}$ is the mass of the scalar field. Furthermore, we will mostly take the scalar $\psi$ to be massless ($m_{\psi} = 0$), unless otherwise indicated.

The above field equations have two scaling symmetries that will turn out to be useful. They are:
\begin{eqnarray}
\label{eq:symmetry1}
& r \rightarrow ar, \quad (t, x, y) \rightarrow (t, x, y) / a, \quad g \rightarrow a^2 g, \quad \phi \rightarrow a\phi, & \nonumber\\
\label{eq:symmetry2}
& e^\chi \rightarrow a^2 e^\chi, \quad t \rightarrow at, \quad \phi \rightarrow \phi / a. &
\end{eqnarray}
We are interested in non-extremal black hole solutions. We use the first symmetry to set the black hole horizon radius $r_{+}$ to 1, and the second symmetry to set $\chi$ to zero at the boundary. With these choices, the following boundary conditions can be used to fix the remaining free parameters in Eq. (\ref{eq:eomstart}):
\begin{eqnarray}
  \label{eq:bcgravity}
  \nonumber \phi(r_{+}) = 0, \qquad \phi'(r_{+}) = E, \qquad g(r_{+}) = 0, \\
  \psi(r_{+}) = \psi_{+}, \qquad \psi'(r_{+}) = \frac{V'(\psi_{+})}{2 \left(3 - \frac12 V(\psi_{+}) - \frac14 E^{2} e^{\chi_{+}} \right) }
\end{eqnarray}
With the above choice of the boundary conditions, the solutions are determined by two parameters: $\psi_+$ and $E$. The latter quantity can be interpreted as the electric field at the horizon. From the equations of motion Eqs. (\ref{eq:eomstart}), the general asymptotic form of the scalar field $\psi$ near the boundary is of the form:
\begin{equation}
 \psi = \frac{\psi_1}r + \frac{\psi_2}{r^2} + \cdots
\end{equation}
The parameters $\psi_+$ and $E$ in Eqs. (\ref{eq:bcgravity}) can be traded for $\psi_1$, $\psi_2$. In what follows we choose to set $\psi_1$ to zero\footnote{We set either $\psi_1$ or $\psi_2$ has to be zero, as we are interested in the field theory in the absence of external sources. Massless scalars allow for two inequivalent quantization, where either
$\psi_1$ or $\psi_2$ is interpreted as a normalizable mode. We choose the quantization for which $\psi_2$ is normalizable; the alternative quantization gives results which are qualitatively similar to the ones discussed below.}. This fixes $E$, and the remaining free parameter $\psi_+$ can be used to vary the temperature of the solution.

\begin{figure}
\begin{center}
\subfigure [Scalar $\psi$]{\label{fig:psiprofile}\includegraphics[scale=0.8]{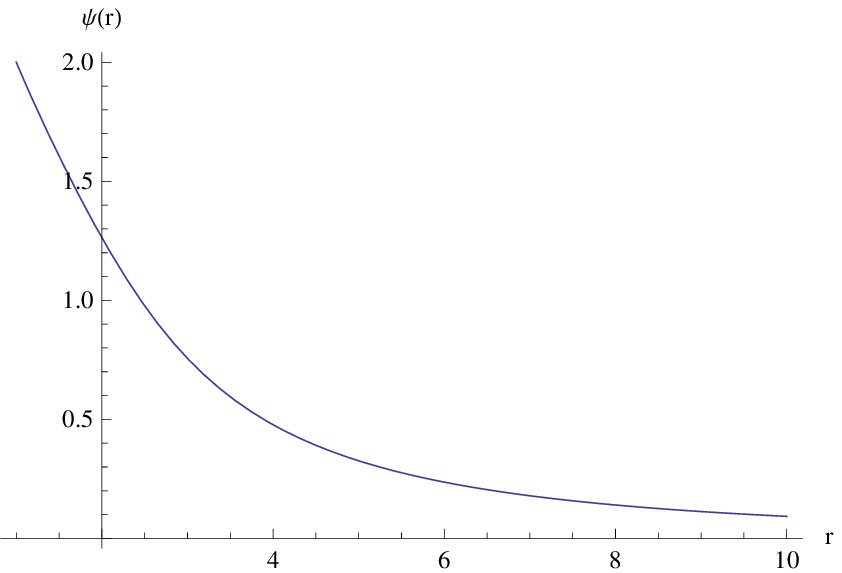}}\hspace{0.5cm}
\subfigure [Metric function $g$]{\label{fig:gprofile} \includegraphics[scale=0.8]{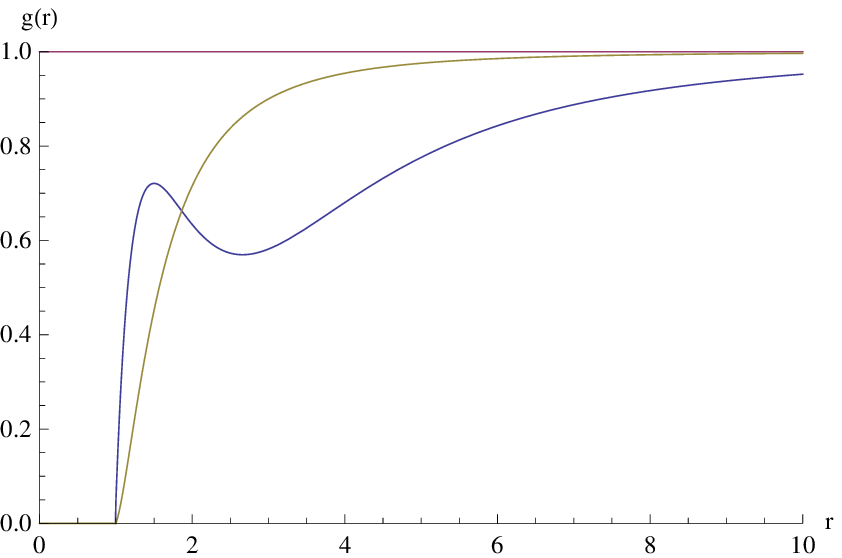}}
 \caption{The profiles for $g(r)$ and $\psi(r)$ at $T_{eff} = 0.036$. For reference, the grey monotonically increasing 
 curve is for the pure RN black hole.}
\label{fig:profiles}
\end{center}
\end{figure}

In Fig \ref{fig:profiles} we plot typical profile for the scalar fied $\psi(r)$ and the function $g (r)$. The temperature is $T_{eff} \equiv \frac{T}{T_{c}}= 0.036$. The expected asymptotic forms for this solution are $\psi \sim 1/r^2$ and $g \sim r^2$ for large $r$.

As the temperature is reduced, the thermodynamics is dominated by one of  two different backgrounds. Above the critical temperature, there is no scalar hair, and the relevant background is the Reissner-Nordstr\"{o}m (RN) black hole. Below the critical temperature, the scalar begins to condense; though around the critical temperature this can be treated as a small perturbation, as we lower the temperature further backreaction ultimately becomes important. Thus we have to use a non-extremal backreacted black hole with scalar hair, constructed in \cite{Hartnoll:2008kx}. As the temperature approaches zero the system resembles the zero temperature backreacted black hole with a zero radius horizon, constructed in \cite{Horowitz:2009ij}. We find that in the zero temperature limit our results approach those of \cite{Faulkner:2009am}, as they should.

\section{Fermions}
\label{sec:fermion}

We now concentrate on the effect of a scalar condensate on non-Fermi liquid behavior. The basic idea, introduced in \cite{Liu:2009dm}, is to introduce fermions in our geometries, whose Green's functions probe the existence and properties of a Fermi surface in the boundary theory. In our case there can be also a scalar field present in the bulk, and therefore we have to specify the coupling between the fermions and the scalar condensate.

We take the action for the bulk fermion $\Psi$ to be:
\begin{equation}
\label{eq:dirac}
S_{\Psi} = \int d^{d+1} x \sqrt{-g}~i(\bar{\Psi}\Gamma^M \mathcal{D}_M \Psi
           - m \bar{\Psi}\Psi -\lambda |\psi|^2\bar{\Psi}\Psi ),
\end{equation}
where $\Gamma^M$ are the curved space Gamma matrices and:
\begin{equation}
\label{eq:diracNotations}
\bar{\Psi} = \Psi^\dagger \Gamma^t, \qquad
\mathcal{D}_M = \partial_M + \frac14 \omega_{abM}\Gamma^{ab} - iq_{f}A_M.
\end{equation}
Here $M$ denotes bulk spacetime indices while $a,b$ denote tangent space indices. Greek letters denote indices along the boundary directions. Thus $\Gamma^{ab}$ are the tangent space Gamma matrices. We also choose units such that $R = 1$ in the $AdS$ geometry. The charge of the fermionic field is denoted by $q_{f}$, and the fermion mass is $m$. We are mostly working with massless fermions, $m=0$.

The last term in Eq.(\ref{eq:dirac}) is a quartic coupling between the scalar $\psi$ and the Dirac fermion $\Psi$, and $\lambda$ is a tunable parameter controlling the coupling to the scalar, assumed to be positive. This  is the most general coupling for general choices of scalar and fermion charges, though cubic couplings can exist when $q_{b}= 2q_{f}$ \cite{Faulkner:2009am}. The quartic coupling can be absorbed into an effective (radial dependent) mass term for the fermions:

\begin{equation}
M(r) \equiv m + \lambda |\psi|^2
\end{equation}

Since the quartic coupling does not necessitate a definite ratio of fermion and boson charge, we are free to vary their ratio. In the following we keep the charge of the complex scalar field to be $q_{b}=1$, which is in the range where a phase transition occurs, but keep the freedom to vary $q_{f}$.

With the quartic coupling only, the mechanics of the fermionic equations remains more or less the same \cite{Liu:2009dm}.  We choose the following basis for the Gamma matrices and the (4 component) spinors:
\begin{equation}
\label{eq:basis}
\Gamma^r = \begin{pmatrix} 1 & 0 \\ 0 & -1 \end{pmatrix}, \quad
\Gamma^\mu = \begin{pmatrix} 0 & \gamma^\mu \\ \gamma^\mu & 0 \end{pmatrix}, \quad
\Psi = \begin{pmatrix} \Psi_+ \\ \Psi_- \end{pmatrix},
\end{equation}
where $\Psi_\pm$ are two-component spinors and $\gamma^\mu$ are (2+1)-dimensional gamma matrices. We can now separate the radial and boundary coordinate dependencies in $\Psi$ as follows:
\begin{equation}
\label{eq:separate}
\Psi_\pm = (-gg^{rr})^{-\frac14}e^{-i\omega t + ik_i x^i} \Phi_\pm, \quad
\Phi_\pm = \begin{pmatrix} y_\pm \\ z_\pm \end{pmatrix}
\end{equation}
For the $\gamma^\mu$, we choose the basis $\gamma^0 = i\sigma_2, \gamma^1 = \sigma_1, \gamma^2 = \sigma_3$. We also use the rotational symmetry of the system to set $k_2 = 0$. The Dirac equations then reduce to two sets of decoupled equations:
\begin{eqnarray}
\label{eq:decoupled}
\sqrt{\frac{g_{ii}}{g_{rr}}} (\partial_r \mp M\sqrt{g_{rr}})y_\pm =
\mp i(k_1 - u)z_\mp, \nonumber\\
\sqrt{\frac{g_{ii}}{g_{rr}}} (\partial_r \pm M\sqrt{g_{rr}})z_\mp =
\pm i(k_1 + u)y_\pm,
\end{eqnarray}
with
\begin{equation}
\label{eq:udef}
u = \sqrt{\frac{g_{ii}}{-g_{tt}}} \left(\omega + q_{f} \phi(r)\right) \quad M = \lambda |\psi|^2
\end{equation}
We define the ratios $\xi_+ = iy_-/z_+$, $\xi_- = -iz_-/y_+$, in terms of which eqs. (\ref{eq:decoupled}) can be written as:
\begin{equation}
\label{eq:xiEquations}
\sqrt{\frac{g_{ii}}{g_{rr}}} \partial_r \xi_\pm = -2M\sqrt{g_{ii}}\xi_\pm
\mp (k_1 \mp u) \pm (k_1 \pm u)\xi_\pm^2.
\end{equation}
The retarded Green's function $G_R$ is given in terms of the quantities $\xi_\pm$ by:
\begin{equation}
\label{eq:retardedgf}
G_R = \lim_{\epsilon \rightarrow 0} \epsilon^{-2M}
\left.\begin{pmatrix} \xi_+ & 0 \\ 0 & \xi_- \end{pmatrix}\right|_{r = \frac1\epsilon}
\equiv \begin{pmatrix} G_{11} & 0 \\ 0 & G_{22} \end{pmatrix}
\end{equation}
The spinors $\xi_{{\pm}}$ satisfy infalling boundary conditions at the black hole horizon, which is located at $r_+ = 1$. This implies:
\begin{equation}
\label{eq:bcfermion}
\lim_{r \rightarrow 1} \xi_\pm(r) = i
\end{equation}
This completes our review of the setup.
We will be interested in the properties of the Fermi surface as we vary the temperature, or the quartic coupling $\lambda$\footnote{Note that the geometry includes back-reaction from the scalar condensate, therefore the condensate influences the fermions even when $\lambda=0$.} In order to do that, we solve numerically the system of equations (\ref{eq:xiEquations}) in the various backgrounds obtained by solving the bosonic equations, Eqs. (\ref{eq:eomstart}). From this, we can calculate the retarded Green's function (Eq. (\ref{eq:retardedgf})) and the fermionic spectral densities (related to the imaginary part of the retarded Green's function), and investigate the effects of turning on $\lambda$ and varying the temperature.

\section{Reissner-Nordstr\"{o}m black hole ($T > T_c$)}
\label{sec:rn}

The Reissner-Nordstr\"{o}m black hole exists for any temperature, and is the dominant phase for $T \geq T_c$. This background was originally discussed in \cite{Liu:2009dm}, and we concentrate here on the behavior at non-zero temperature, to set up our notation and provide a comparison to other backgrounds. The Reissner-Nordstr\"{o}m metric is given by:
\begin{equation}
\label{eq:rn}
ds^2 = -f(r) \, dt^2 + \frac{dr^2}{f(r)} + r^2 (dx^2 + dy^2)
\end{equation}
with $f(r) = r^2 + \frac{Q^2}{r^2} - \frac{1 + Q^2}{r}$, $\phi(r) = \mu \left(1 - \frac1r \right)$ and $\mu = Q$. Here we have used the scaling symmetries in Eqs. (\ref{eq:symmetry1}) to  set $L=1$ and horizon $r_+=1$, then the temperature is
\ba
T=\frac{1}{4\pi}(3-Q^2),
\ea
and the dimensionless temperature is
\ba
T_{eff} \equiv \frac{T}{\mu}=\frac{1}{4\pi}\frac{(3-Q^2)}{Q}.
\ea
Since the scalar $\psi=0$, the black hole charge $Q$ is the only tunable parameter.

The discussion in \cite{Liu:2009dm} concentrated on the spectral density of the fermions at zero temperature, finding a signal of a Fermi surface by the presence of a delta function peak in the spectral function (or equivalently, a pole in the Green's function), at critical value of the momentum ($k=k_{F}$) and zero frequency. As expected, we find that at finite temperature, the pole in the Green function moves off the real axis, to complex values of $\omega$. In the following, we investigate how the pole evolves in the complex $\omega$ plane as the momentum $k$ changes.

\begin{figure}[h!]
\begin{center}
\includegraphics[scale=0.8]{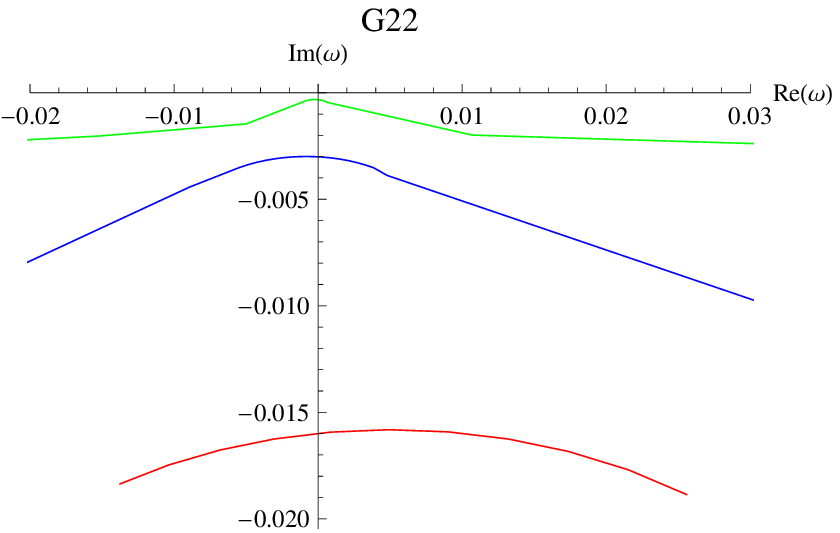}
\includegraphics[scale=0.8]{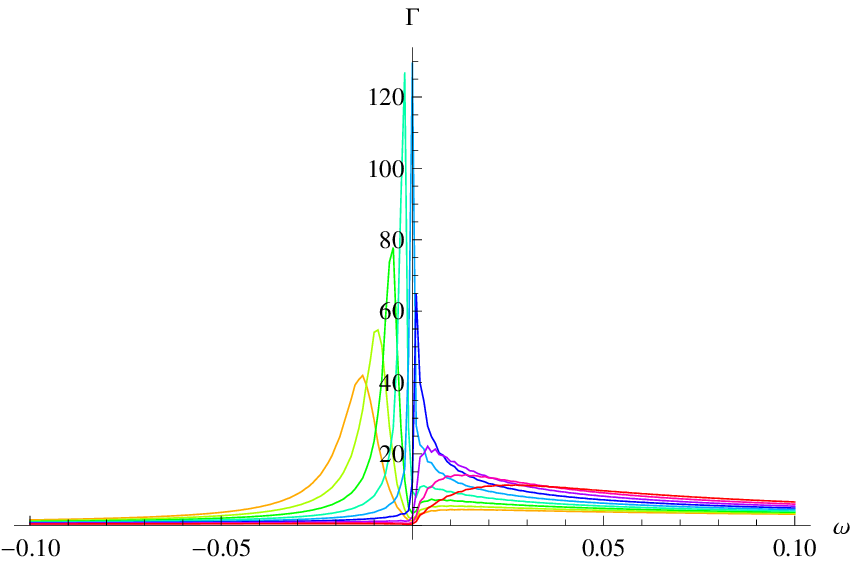}
\caption{The left figure shows the $G_{22}$ pole location in the complex frequency plane as function of momentum, at different temperatures,
$T_{eff}\approx (0.00016,0.0016,0.008)$ respectively from top to bottom.
The right figure shows a plot of $\mathrm{Im} \, G_{22}$ vs. $\mathrm{Re} \, \omega$ at $T_{eff}\approx 0.00016$ for a range of values of $k$: from $0.8$ to $1$ for the curves from yellow to red. The curve with highest ${\mathrm{Im} \, G_{22}}$ has $k=0.9$.
}
\label{fig:peaks_RN_Tvar}
\end{center}
\end{figure}

\noindent In Fig \ref{fig:peaks_RN_Tvar} we plot the typical trajectory of the pole of the spectral function ($\mathrm{Im}(G_{{22}})$) on the complex $\omega$ plane for a range of temperatures, in the left panel. The position of the pole changes as the momentum $k$ is varied. The Fermi momentum corresponds to the point of closest approach to the real axis. The right panel shows the frequency dependence of the spectral function for one choice of the temperature.

\begin{figure}[h!]
\begin{center}
\subfigure [$\Gamma$ vs. temperature] {\label{fig:gtemp}\includegraphics[scale=0.8]{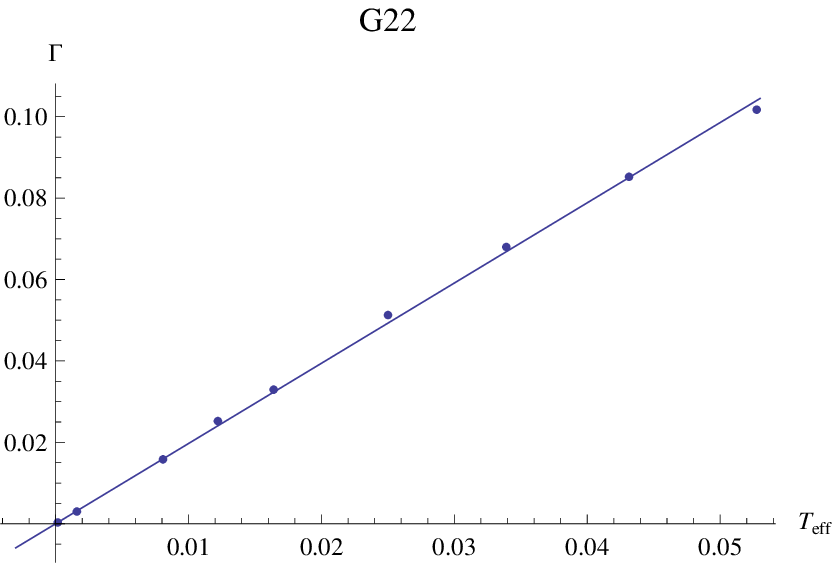}}\hspace{0.5cm}
\subfigure [Pole position vs. momentum]{\label{fig:gmom} \includegraphics[scale=0.8]{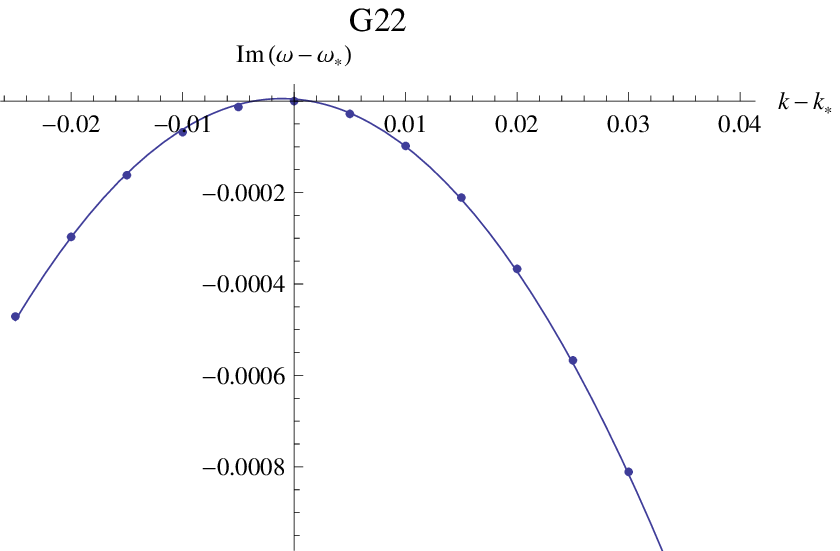}}
\end{center}
 \caption{Variation of the pole position with temperature and momentum. The plots are for $G_{22}$.}
\label{fig:gexp}
\end{figure}

From this data we can determine how the minimum distance of the pole $\Gamma$ from the real axis changes with temperature\footnote{In all the plots of the poles in the complex frequency plane, we concentrate on the vicinity of primary Fermi surface, which is the range of momenta where the poles of the spectral function are closest to the real axis. Other branches of the curves shown exist in some cases.}. We find that $\Gamma$ grows linearly with temperature (Fig \ref{fig:gtemp}). This is consistent with the results of \cite{Liu:2009dm}, and extends them beyond the low temperature regime.

We can also measure how the imaginary part of the pole position changes as a function of $k - k_F$, near the Fermi momentum. For the RN black hole at finite temperature, $\Gamma \sim (k - k_F)^z$, with $z \sim 2$. The critical exponent $z$ does not depend strongly on the temperature, and is close to the value of the exponent found at  \cite{Liu:2009dm} for $T=0$. The dependence of $\Gamma$ on the temperature and on $k - k_F$ is summarized in Fig \ref{fig:gmom}.

\section{Non Extremal Hairy Black Hole  ($T < T_c$)}
\label{sec:backreacted}

We now consider the backreaction of the scalar on the black hole geometry, at temperatures below the critical one. The fermions are still treated as probes. At first we do not include any coupling between the scalar and fermions, concentrating on the behavior as a function of temperature and the fermion charge. The influence of the scalar on the behavior of the fermions comes through the backreaction on the geometry. We then turn on the quartic coupling $\lambda$ discussed above, and demonstrate the influence of changing the quartic coupling, the fermion charge and the temperature on features of the spectral functions of the fermions.

\begin{figure}[h!]
\begin{center}
\includegraphics[scale=0.8]{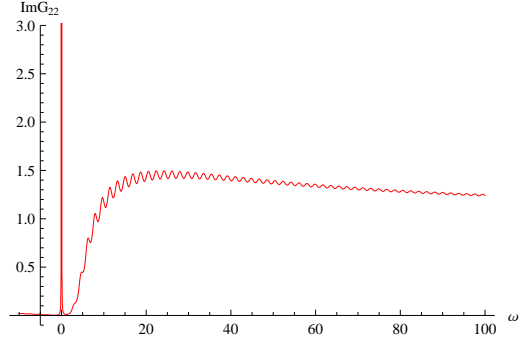}
\caption{$\mathrm{Im}(G_{22})$ vs. frequency $\omega$ at the Fermi momentum. This is the so-called ``peak-dip-hump" behaviour, with some ripples in the large $\omega$ region. The data is for $T_{eff}=0.004,~q_F=1,\lambda=0$.
}
\label{fig:kF_realomega}
\end{center}
\end{figure}

Fig. \ref{fig:kF_realomega} shows a plot of $\mathrm{Im}(G_{22})$ vs. frequency $\omega$ for zero coupling. We see the typical ``peak-dip-hump" found in \cite{Chen:2009p10965}.

\begin{figure}[h!]
\begin{center}
\includegraphics[scale=0.8]{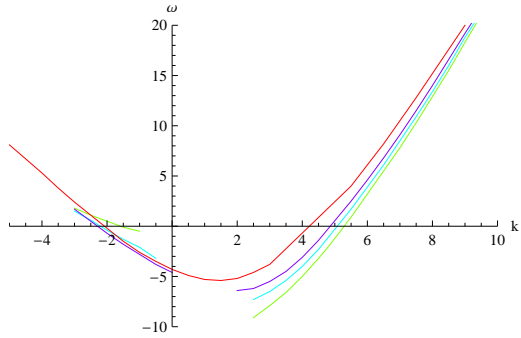}
\caption{Peak position ($k$, $\omega$) as a function of coupling. $\lambda={0,0.5,1,2}$ from bottom to top (green to red).
}
\label{fig:lambda_var_realomega}
\end{center}
\end{figure}

In Fig. \ref{fig:lambda_var_realomega} we plot the location of the pole in the spectral density $\mathrm{Im}G_{22}$ as function of momentum and frequency, for various values of the coupling, including $\lambda=0$. Note that unlike the high temperature case, discussed in the previous section, the spectral density  $\mathrm{Im}G_{11}$ exhibits some non-analytic behavior, though for any momentum and frequency the peak of $\mathrm{Im}G_{11}$ is much broader than the one of $\mathrm{Im}G_{22}$. For that reason we concentrate exclusively on the spectral density $\mathrm{Im}G_{22}$ below.

The qualitative features of Fig. \ref{fig:lambda_var_realomega} correspond to the existence of a fairly sharp Fermi surface (smoothed out by finite temperature), with gapless fermionic excitations. Unlike the high temperature case, and similar to the discussion in  \cite{Faulkner:2009am} (for zero temperature), we find a line of sharp excitations in the frequency-momentum plane, rather than an isolated point. Note that we find sharp excitations for both signs of the momentum, or in other words both for particle and hole excitations.

\begin{figure}[h!]
\begin{center}
\includegraphics[scale=0.8]{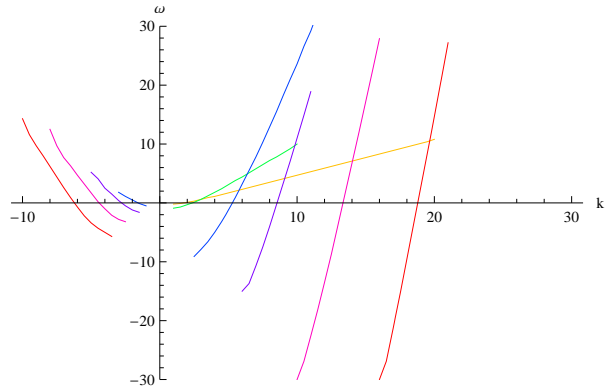}
\caption{Pole position ($k$, $\omega$) as a function of temperature. The temperature $T_{eff} = 0.023, 0.014, 0.004, 0.002, 0.0013, 0.0009$ from yellow to red (left to right). The other parameters are $q_F = 1$, $\lambda = 0$.}
\label{fig:T_var_realomega}
\end{center}
\end{figure}

Fig. \ref{fig:T_var_realomega} is similar to Fig. \ref{fig:lambda_var_realomega}, except that here the temperature is varied for a fixed coupling. We can also determine the behavior of the imaginary part of the pole position in the complex $\omega$ plane as a function of temperature, similar to what we did in Sec. \ref{sec:rn}. We find that in this case the temperature dependence is no longer linear, rather we find that  for $\lambda=0$, $\mathrm{Im}~\omega_p \sim T^{\eta}$, with $\eta\sim 5$.

Turning on and varying the scalar-fermion coupling (in addition to the fermion charge) corresponds to scanning different boundary theories, as opposed to variation of physical parameters such as the temperature.  In all those boundary theories, we find that the qualitative properties of the system do not change much: in the condensed phase we still have stable fermionic excitations (as zero temperature) which are gapless\footnote{These observations are consistent with  the expectations in \cite{Faulkner:2009am}.}. It is surprising to find such stable excitations for a large set of theories, particularly for both signs of the coupling $\lambda$. Perhaps this is somehow related to another feature that sets these theories apart, namely having a dual description in terms of classical gravity.

\begin{figure}[h!]
\begin{center}
\includegraphics[scale=0.4, angle=270]{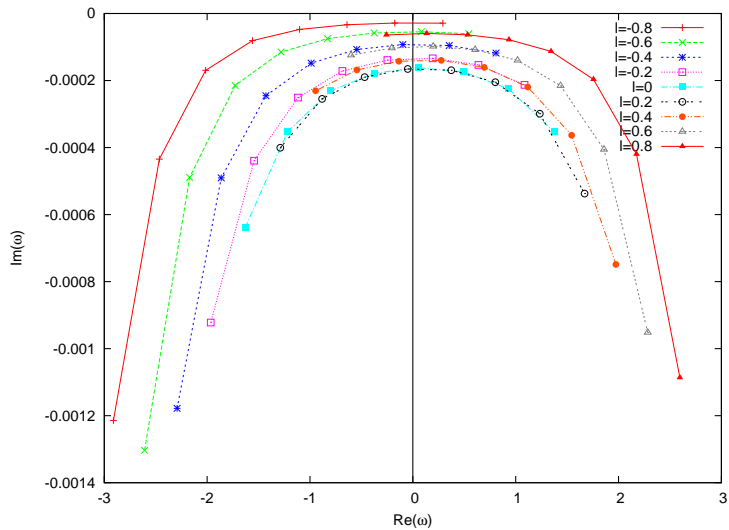}
\end{center}
\caption{Contour plots for varying $\lambda$ for temperature $T_{eff} \sim 0.004$.}
\label{fig:lambda}
\end{figure}

In Fig \ref{fig:lambda} we plot the contour of poles for changing $\lambda$, at a fixed temperature. The coupling $\lambda$ here determines the range of values of $k$ which give rise to poles, and there are no gapped excitations in this picture. In the zero temperature limit , the gapless excitations become stable.
We find that turning on $\lambda$ (for a fixed temperature) seems to stabilize the fermionic excitations, for either sign of $\lambda$.  For example, the critical exponent $z$ defined above and in \cite{Liu:2009dm} increases monotonically with $|\lambda|$.

\begin{figure}[h]
\begin{center}
%\subfigure [Im_omega vs. fermion's charge] {{\label{fig:Gamma_qF}
\includegraphics[scale=0.8]{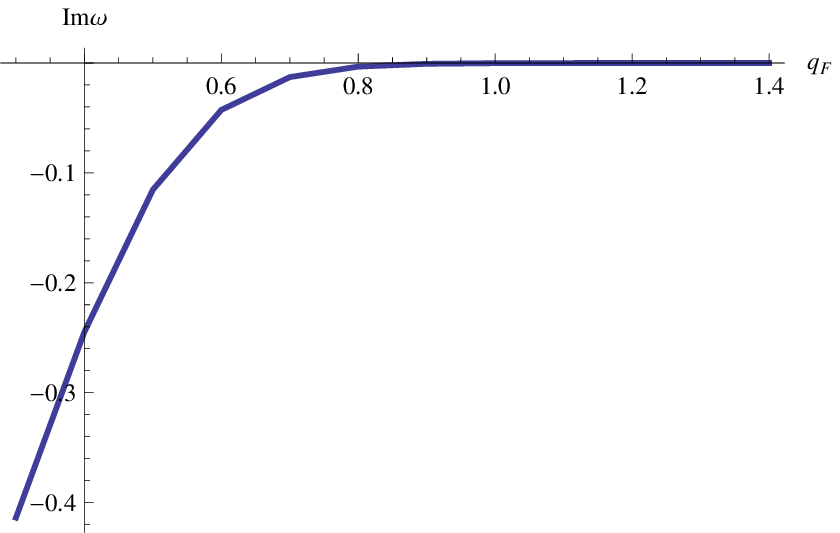}
%}
\hspace{0.5cm}
%\subfigure [kF vs. fermion's charge]{\label{fig:kF_qF}
\includegraphics[scale=0.8]{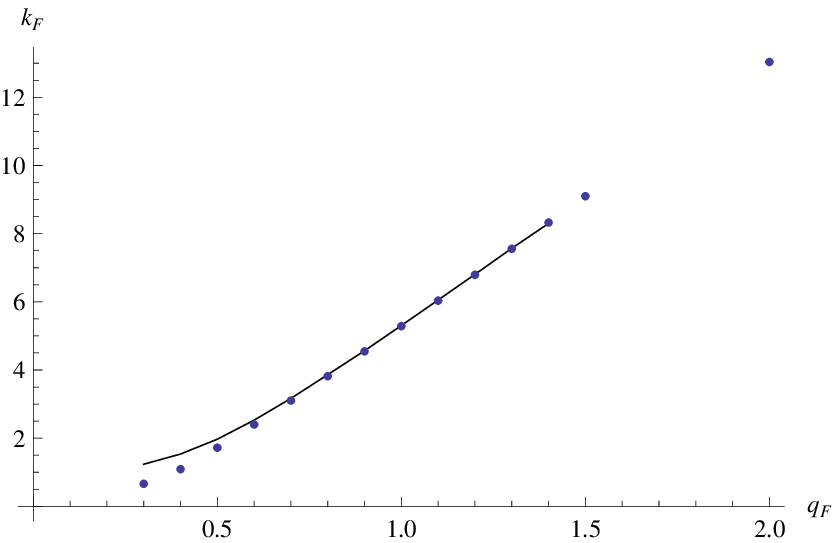}
%}
\end{center}
 \caption{The pole position as function of fermion charge ${q_{F}}$. The plots are for $T_{eff}=0.004$. As ${q_{F}}$ increases, the curve in the complex $\omega$ plane moves
 closer to the real axis, and the fermion excitations become more stable. In the left panel, $\mathrm{Im}(\omega)$ is plotted as function as $q_{F}$, while the right panel shows the corresponding Fermi momenta $k_F$.}
\label{fig:qF}
\end{figure}

We also checked the dependence of these feature on the fermion charge $q_{F}$. These results are summarized in figure \ref{fig:qF}. In this case as well, we find that while qualitative features do not change as the fermion charge is varied, the detailed features do.

\bigskip
\bigskip
\bigskip

\section*{Acknowledgements}

We thank Clifford Johnson, Matt Roberts and Mark van Raamsdonk for useful conversations and correspondence. The work is supported by discovery grant from NSERC.

\newpage

\bibliography{fermi_nonextremal.bib}

\end{document}